# Simulation of the Magnetization Dynamics of a Single Domain BiFeO$_3$ Thin Film

Yu-Ching Liao[1], Dmitri E. Nikonov[2], Sourav Dutta[1,3], Sou-Chi Chang[2], Sasikanth Manipatruni[2], Ian A. Young[2], and Azad Naeemi[1]

[1]School of Electrical and Computer Engineering, Georgia Institute of Technology, Atlanta, GA 30332 USA

[2]Components Research, Technology & Manufacturing Group Intel Corporation, Hillsboro, OR 97124 USA

[3]Department of Electrical Engineering, University of Notre Dame, Notre Dame, IN 46556, USA

**Abstract:**

The switching dynamics of a single-domain BiFeO$_3$ thin film is investigated through combining the dynamics of polarization and Neel vector. The evolution of the ferroelectric polarization is described by the Landau-Khalatnikov (LK) equation, and the Landau–Lifshitz–Gilbert (LLG) equations for spins in two sublattices to model the time evolution of the antiferromagnetic order (Neel vector) in a G-type antiferromagnet. This work theoretically demonstrates that due to the rotation of the magnetic hard axis following the polarization reversal, the Neel vector can be switched by 180$^0$, while the weak magnetization can remain unchanged. The simulation results are consistent with the ab initio calculation[1], where the Neel vector rotates during polarization rotation, and also match our calculation of the dynamics of order parameter using Landau-Ginzburg theory[2]. We also find that the switching time of the Neel vector is determined by the speed polarization switching and is predicted to be as short as 30 ps.

Introduction

Recently, the demand for low-power non-volatile devices has led a growing interest in beyond-CMOS spintronic devices. Among them, magnetoelectric (ME) devices are promising as they are voltage-driven and dissipate less energy compared to current-driven spintronic devices[3,4]. However, the speed of the proposed ME devices are slower compared to the high performance CMOS[5,6] transistors. Multiferroic materials showing at least two of the possible simultaneously occurring characteristics- ferroelectricity, ferromagnetism, antiferromagnetism, and ferroelasticity[7] are promising for ME devices due to their intrinsic magnetoelectric coupling. In addition, ME devices using the exchange interaction between a ferromagnet and a multiferroic material exhibit deterministic 180⁰ magnetization switching rather than 90⁰ easy axis rotation typical for piezoelectric-magnetostrictive heterostructures. Recent calculations show that an antiferromagnet can switch on the order of 10 ps[8,9], much faster than the ferromagnet because of the strong exchange coupling and the characteristic of inertia-driven switching[10]. Bismuth ferrite, BiFeO$_3$, possessing ferroelectricity, G-type antiferromagnetism, and weak magnetization, is the only single-phase multiferroic material known so far, that has both the Curie temperature ($T_c$~1103K) and the Neel temperature ($T_N$~643K) well above room temperature, which makes it a promising candidate for the room-temperature application. Currently, most experimental demonstrations[1,11,12] and theoretical simulations[13,14] regarding $BiFeO_3$ focus on the magnetic textures of $BiFeO_3$ under high magnetic field[2] or epitaxial strain[15], the polarization switching time[16], or the multidomain switching of $BiFeO_3$/CoFe heterojunction[12,13]. However, there are no comprehensive studies of the spin switching in $BiFeO_3$ considering its application in ultra-scaled beyond CMOS devices. Therefore, in this paper, we aim at analyzing the switching dynamics and the switching time of the magnetic order in a single-domain $BiFeO_3$ thin film after applying the electric field. We solve the magnetic order of $BiFeO_3$ by using the Landau-Ginzburg phenomenological theory to derive both LK and LLG equations. Our simulations of the magnetic dynamics show that the Neel vector switches $180^0$ after the polarization reversal. We also verify the consistency of the results with those obtained via solving the effective equation of motion for the Neel vector.

Bulk $BiFeO_3$ has a rhombohedrally distorted perovskite structure and belongs to the space symmetry group R3c, while the crystal structure and space group in a thin film $BiFeO_3$ may vary from tetragonal to orthorhombic depending on the compressive or tensile strain[17]. The ferroelectricity of $BiFeO_3$ mostly originates from the displacement of $Bi^{3+}$ ions relative to the rest of the lattice under an applied electric field. Thus, the application of a sufficiently strong applied electric field will reverse the polarization[11,18]. Interestingly, during polarization reversal, the iron ions and oxygen octahedra also rotate, which causes the weak magnetization to rotate[1]. The weak magnetization in $BiFeO_3$ originates from the Dzyaloshinskii-Moriya interaction (DMI)[19,20], which is due to the tilting of oxygen octahedra from the ideal $ABO_3$ perovskite structure combined with the spin-orbit coupling effect. The Hamiltonian of DMI is expressed as

$$E_{DMI} = \sum_{i=1}^{N} \sum_{j \neq i}^{nn} \boldsymbol{D}_{i,j} \cdot (\boldsymbol{S}_i \times \boldsymbol{S}_j), \qquad (1)$$

where $\boldsymbol{D}_{i,j}$ is the DM vector between cells $i$ and $j$, and $\boldsymbol{S}_i$ and $\boldsymbol{S}_j$ are the spin vectors of cells $i$ and $j$, respectively. The direction of the DM vector is determined by the cross product of the displacement of oxygen ion $\boldsymbol{x}$ from the midpoint between Fe ions and the distances between two Fe ions $\boldsymbol{r}_{ij}$ so that $\boldsymbol{D}_{i,j} = V_0(\boldsymbol{r}_i \times \boldsymbol{r}_j) = V_0(\boldsymbol{r}_{ij} \times \boldsymbol{x})$ where $V_0$ is the microscopic constant. From a crystallographic point of view, the rotation of the oxygen octahedra is anti-phase when viewing from the rotation axis, which is also called anti-ferrodistortive axis $\boldsymbol{AFD}$, of $BiFeO_3$ [21]. Hence, the displacement of oxygen ion, $\boldsymbol{x}$, and the DM vector $\boldsymbol{D}_{i,j}$ are both staggered vectors or quasi-axial vectors, and the direction of $\boldsymbol{D}_{i,j}$ is determined by the $\boldsymbol{AFD}$.[21] Ab initio calculations[1] have shown that $\boldsymbol{D}_{i,j}$ is nearly parallel to the polarization during polarization rotation so that this ME coupling will determine the switching dynamics in our model. One can also re-write the Hamiltonian of DMI as

$$E_{DMI} \approx \sum_{i=1}^{N} \boldsymbol{D}_{i,j} \cdot (\boldsymbol{N} \times \boldsymbol{M}_c) = \boldsymbol{H}_{DMI} \cdot \sum \boldsymbol{S}_i^{AFM} \qquad (2)$$

considering there are two sublattices $A$ and $B$ where $\boldsymbol{N} = \sum_{i,j}(\boldsymbol{S}_i^A - \boldsymbol{S}_j^B)$ is the Neel vector, $\boldsymbol{M}_c = \sum_{i,j}(\boldsymbol{S}_i^A + \boldsymbol{S}_j^B)$ is the weak-canted magnetization, and $\boldsymbol{H}_{DMI} = \boldsymbol{D}_{i,j} \times \boldsymbol{N}$ is the effective magnetic field of DMI since the direction of $S_i \times S_j$ is equal to $N \times M_c$. Note that because of the cross-product relation between $D_{i,j}$, $N$, and $M_c$, these three vectors always form a right-handed system[22]. We will later compare the results based on these two expressions of the DMI field to verify the equivalency.

**Ferroelectric and magnetic dynamics simulation model**

We aim to simulate the spin dynamics and the switching time of a single-domain area of BiFeO$_3$ thin film whose magnetization can be controlled by an external electric field. We focus on a single-domain BiFeO$_3$ thin film because of its potential application in beyond-CMOS devices scaled to nanometer sizes. Previous studies[1] have shown that the Bi$^{3+}$ ion displacement, Fe ion displacement, and the tilting angles of oxygen octahedra have similar switching paths, which indicates that the ferroelectricity and weak ferromagnetism are coupled during switching, and the switching of the magnetization in BiFeO$_3$ can be determined by combining the ferroelectric and micromagnetic models. In other words, the axial vector $D_{i,j}$ in the micromagnetic simulation can be modified by the polarization in each time step in the ferroelectric dynamics. Therefore, in this work, we simulate the dynamics of $N$, and $M_c$ in a single-domain BiFeO$_3$ thin film following the rotation of polarization and $D_{i,j}$ as shown in Figure 1.

The polarization ($P$) dynamics of BiFeO$_3$ is described by the LK equation

$$\gamma_{FE} \frac{\partial P_i}{\partial t} = -\frac{\partial F}{\partial P_i}, \quad (3)$$

where $\gamma_{FE}$ is the 'viscosity coefficient', and $F$ is the total free energy of the ferroelectric. The simplified total free energy of the single-domain ferroelectric is expressed as[23]:

$$F(\boldsymbol{P}, \boldsymbol{\varepsilon}) = \alpha_1(P_1^2 + P_2^2 + P_3^2) + \alpha_{11}(P_1^4 + P_2^4 + P_3^4) + \alpha_{12}(P_1^2 P_2^2 + P_1^2 P_3^2 + P_2^2 P_3^2) + K_{Strain}(\boldsymbol{P} \cdot \boldsymbol{u})^2 - (c\frac{(P_{dw}-P)\cdot P}{\epsilon_r \epsilon_0} + \boldsymbol{P} \cdot \boldsymbol{E}_{ext}), \quad (4)$$

where $\alpha_1, \alpha_{11}, \alpha_{12}$ are the phenomenological Landau expansion coefficients, $\boldsymbol{u}$ is the axis of substrate strain, $K_{Strain}$ is the strain energy, $c$ is the geometry factor of the averaged domain wall, $\boldsymbol{P}_1, \boldsymbol{P}_2$, and $\boldsymbol{P}_3$ are the polarization components in *x*, *y*, and *z*- directions, $\boldsymbol{E}_{ext}$ is the external electric field. We take the case of intrinsic strain due to deposition to result in an easy plane normal to the [0 1 1][1]. The last term in (4) represents the depolarization energy and the external field energy, and $\boldsymbol{P}_{dw}$ is the polarization in an adjacent domain. The depolarization energy is calculated by approximating the depolarization field due to bound charges as the difference between polarizations on the sides of the domain wall. Note that the ferroelectric domain wall here is considered as the averaged contribution to the single domain ferroelectric. We then assume an input pulse of the electric field with the duration of 80 ps and a rise time of 5 ps. With these assumptions, the trajectory of polarization switching, i.e. the two-steps polarization switching, is in agreement with the ab initio calculation[1] as will be shown later. The parameters of the ferroelectric model are listed in Table 1.

We start by using the Landau-Ginzburg (LG) theory to simulate the spin dynamics. Landau-Ginzburg theory[15,2,24] usually describes the antiferromagnetic order of BiFeO$_3$ by a Neel vector $\bm{n}$. Similar to the spin Hamiltonian discussed previously, $\bm{n}$ and $\bm{m}$ are microscopically defined as the vectors of the difference and the summation of the magnetic moment densities of the two sublattices in an antiferromagnet, respectively. The total free energy density of a BiFeO$_3$ thin film includes the homogeneous and inhomogeneous exchange energies $F_{exch} = a\bm{m}^2 + A\sum(\nabla \bm{n}_i)^2$, magnetic anisotropy energy $F_{an} = K_{AFM}(\bm{e}_p \cdot \bm{n})^2$, magnetoelastic energy $F_{M\_elast} = K_{epi}(\bm{e}_s \cdot \bm{n})^2$, and DMI energy $F_{DMI} = X_\perp \bm{H}_{DMI}^2 \frac{(\bm{e}_p \cdot \bm{n})^2}{2}$ where $a$ is the homogeneous exchange constant, $A$ is the inhomogeneous exchange constant, $K_{AFM}$ is the effective anisotropy energy which includes single-ion anisotropy energy and effective anisotropy energy that comes from DM interaction, $\bm{e}_p$ is the hard axis of the antiferromagnet, $K_{epi}$ is the anisotropy energy originates from epitaxial constraint, $\bm{e}_s$ is the axis normal to the film, $X_\perp$ is the magnetic susceptibility that is perpendicular to $\bm{n}$, and $\bm{H}_{DMI}$ is the DM field. In our model, we assume the single-domain BiFeO$_3$ thin film has a homogeneous weak magnetization without spin cycloid such that $F_{exch}$ vanishes; thus, only $F_{an}, F_{M\_elast}$ and $F_{DMI}$ are involved. By taking the derivative of magnetic free energy with respect to $\bm{n}$, we get the Neel field as:

$$\bm{f}_n \equiv -\delta_n F = -X_\perp H_D^2(\bm{e}_p \cdot \bm{n})\bm{e}_p - 2K_{AFM}(\bm{e}_p \cdot \bm{n})\bm{e}_p - 2K_{epi}(\bm{e}_s \cdot \bm{n})\bm{e}_s, \qquad (5)$$

The dynamics of $\bm{n}$ is then solved by the effective equation of motion for the Neel field[25]

$$\frac{\ddot{\bm{n}}}{\tilde{\gamma}} = G_1 \bm{f}_n + a[\gamma \bm{f}_n - G_2 \dot{\bm{n}}], \qquad (6)$$

where $\tilde{\gamma}$ is the effective gyromagnetic ratio, $G_1$ and $G_2$ are the phenomenological Gilbert damping parameters.

To model both $N$ and $M_c$ of the antiferromagnet more accurately, we also use the micromagnetic simulation. Previous studies already demonstrated that the dynamics of an antiferromagnet using the Neel vector is the same as the dynamics of a micromagnet with two sublattices[26] where they both conclude the same equation of motion of the Neel vector. Therefore, we further describe a G-type antiferromagnet with two sublattices $a$ and $b$, where the magnetization of one sublattice is antiparallel to another sublattice and the dynamics is solved separately for each cell. We will later calculate the time evolution of the order parameter $\bm{n}$ to compare with the results solved by a micromagnet with two sublattices. The dynamics of a micromagnet is solved by the LLG equation in each cell and the time-varying polarization is assumed homogeneously distributed in the sample. We use both the Object Oriented MicroMagnetic Framework (OOMMF)[27] and our own numerical micromagnetic model based on the finite difference method to account for the microscale interactions such as exchange coupling and compare the results of the two different expressions of DMI. The dynamics of an antiferromagnet is obtained by calculating

$$\dot{\bm{M}}_i = -\gamma(\bm{M}_i \times \bm{H}_i) + \frac{\alpha_G}{M_s}(\bm{M}_i \times \dot{\bm{M}}_i), \qquad (7)$$

where $\bm{M}_i$ is the magnetization in cell $i$, $\bm{H}_i$ is the effective local magnetic field in cell $i$, $M_s$ is the saturation magnetization, and $\alpha_G$ is the Gilbert damping factor. In our model, $\bm{H}_i = \bm{H}_{ex,ij} + \bm{H}_{ani} + \bm{H}_{dem,i} + \bm{H}_{DMI}$ where $\bm{H}_{ex,ij}$ is the exchange coupling field from cell $i$ to its six nearest neighboring cells $j$, $\boldsymbol{H}_{ani}$ is the anisotropy field composed of bulk anisotropy field from DMI and the anisotropy field that comes from compressive epitaxial constraint, $\boldsymbol{H}_{dem,i}$ is the demagnetization field in cell $i$, and $\boldsymbol{H}_{DMI}$ is a magnetic field arising from DM interaction in our OOMMF model whereas in our own micromagnetic solver, the DM Hamiltonian is used. We will compare the differences in the later discussion. The numerical micromagnetic model considers the discretization of the field with each term in the effective field expressed as:

$$\boldsymbol{H}_{ex,ij} = \frac{2A_{AFM}}{\mu_0 M_s} \sum \frac{\boldsymbol{m}(r_i-\Delta_j) - 2\boldsymbol{m}(r_i) + \boldsymbol{m}(r_i+\Delta_j)}{|\Delta_j^2|}, \tag{8}$$

$$\boldsymbol{H}_{ani} = \boldsymbol{H}_{ani,bulk} + \boldsymbol{H}_{ani,epi} = \frac{2K_{AFM}}{\mu_0 M_s}(\boldsymbol{m} \cdot \boldsymbol{P}) \cdot \boldsymbol{P} + \frac{2K_{EPI}}{\mu_0 M_s} m_z, \tag{9}$$

$$\boldsymbol{H}_{DMI} = -\frac{1}{\mu_0 M_s} \frac{\partial E_{DMI}}{\partial \boldsymbol{m}_i} = -\frac{1}{\mu_0 M_s} \boldsymbol{M}_j \times \boldsymbol{D}_{i,j}, \tag{10}$$

$$\boldsymbol{H}_{dem,i} = \frac{-1}{4\pi} \int \frac{n \cdot M(r')(r-r')}{|r-r'|^3} d^2 r' \tag{11}$$

integrating all the other cells besides cell $i$. The details of $\boldsymbol{H}_{dem,i}$ calculation are explained in [28]. Note that the polarization direction determines the hard axis of the bulk anisotropy field because $\boldsymbol{D}_{i,j}$ is a uniaxial vector that is parallel to polarization ($\boldsymbol{P}$)[29], and the hard axis of compressive epitaxial constraint anisotropy field is along [0 0 1][15,24]. The schematic of the rotation of easy plane state during polarization reversal is shown in the experiment[29].

The simulation parameters in the micromagnetic model are listed in Table 2 [13]. The negative sign of $K_{AFM}$ and $K_{EPI}$ refers to the magnetic easy plane state. The sample size of BiFeO$_3$ is 32 nm thick and 20 nm wide and long, and the mesh size is $5 \times 5 \times 1\ nm^3$.

**Results and discussion**

We analyze the switching dynamics of BiFeO$_3$ by introducing two unitless vector variables $\boldsymbol{N} = (\boldsymbol{M}_1 - \boldsymbol{M}_2)/(|\boldsymbol{M}_1| + |\boldsymbol{M}_2|)$ and $\boldsymbol{M}_c = (\boldsymbol{M}_1 + \boldsymbol{M}_2)/(|\boldsymbol{M}_1| + |\boldsymbol{M}_2|)$, which refer to the Neel vector and the weak-canted magnetization of the antiferromagnet, respectively. At the initial stage before the switching starts, Figure 2(a) shows that $\boldsymbol{M}_c$, which is calculated by averaging the magnetization of in every 1 nm of the thickness, is staggered in the *x*, *y*, and *z* directions particularly close to the surface. To understand the staggered behavior of $\boldsymbol{M}_c$ close to the surface, note that within the bulk the anisotropy field and DM field are small compared to the antiferromagnetic exchange coupling field as each cell is surrounded by six nearest neighbors. The exchange coupling field for the surface cells, however, is caused by fewer neighboring cells. Because of the weaker exchange field, the spin vectors on the surface tilt more by the DM field compared to the cells in the bulk region as shown in Figure 2(b). However, since the exchange coupling field is a short-range order field, which becomes negligible to the second nearest neighbors, the staggered $\boldsymbol{M}_c$ only occurs at the surface of the BiFeO$_3$. Conversely, the $\boldsymbol{M}_c$ shows a net magnetization in the bulk region because the spin vectors are almost antiparallel to each other and the DM field creates a weak, unidirectional magnetization. The magnitude of DM field is calculated by comparing the value of $\boldsymbol{M}_c$ with the saturation magnetization in the hysteresis loop of the BiFeO$_3$ thin film in experiment[30]. The peak value of $M_c$ is approximately $1.8 \times 10^3$ A/m, which corresponds to about 1000 Oe $H_{DMI}$ in our model in Figure 3.

Next, to understand the energy barrier and the preferred axis during polarization reversal, we plot the energy landscape considering bulk anisotropy energy and anisotropy energy that comes from epitaxial constraint in Figure 4. We define the polar angle between magnetization and *x*-axis as $\theta$; and the azimuthal angle between magnetization and the *y*-axis on the *yz* plane as $\varphi$. The results show that when $H_{ani,epi}$ is larger than $H_{ani,bulk}$, the lowest energy of magnetization lies on the *xy* plane, and the maximum energy barrier happens at [0 0 1] and [0 0 -1]. However, when $H_{ani,bulk}$ is larger than $H_{ani,epi}$, the preferred easy plane changes to [1 -1 1], where the hard axis is parallel to the polarization direction. Considering the superposition of $H_{ani,bulk}$ and $H_{ani,epi}$, the preferred easy axis becomes along [1 1 0], which is consistent with the experiment results[29]. This is because the epitaxial constraint lifts the preferred easy plane state and results in a unique easy axis state. Because of the rotation of the $H_{ani,bulk}$ along with the polarization, the easy axis of BiFeO$_3$ rotates during polarization switching. Figure 4 also show that the energy barrier of magnetization switching in BiFeO$_3$ is small since the single-ion anisotropy energy from Fe is small when it is half-filled in *d* orbitals[31].

We now study dynamics of **P**, **N** and $M_c$ in BiFeO$_3$ by applying a negative electric field $E_{ext} = 3 \times 10^8$ $A/m$. Figure 5 demonstrates that the switching curves of **P** matches qualitatively to the ab initio calculation in [1], and **N** switches $180^0$ while $M_c$ remains in the same direction after the polarization switching. The switching curves of **N** also show that **N** would first rotate $90^0$ to [-1 1 0] and then rotate to [1 1 0], which is consistent to the ab initio calculation results in [1]. Next, we find that both the LG theory, which solved the order parameter **n**, and the OOMMF model, which solved for both **N** and $M_c$, show the same results that **N** switches $180^0$ during polarization reversal in Figure 5. This demonstrates that our micromagnetic model is consistent with the previous approaches that used LG theory[2,15]. However, unlike LG theory, our micromagnetic model can describe canted magnetization ($M_c$). Note that the final **N** is not precisely along [1 1 0] because in a BiFeO$_3$ thin film, the polarization will deviate from ideal [1 -1 1] in tetragonal lattice. Hence, **N**, which is perpendicular to **P**, deviates from [1 1 0]. For $M_c$, the *x* and *y* components are oscillating to -*x* and +*y* directions initially. However, because of the right-handed relation governing **D**, **N**, and $M_c$, $M_c$ rotates back to the initial direction. Therefore, both **D** and **N** switch $180^0$ after polarization reversal while $M_c$ remains non-switched. Note that the driving force for magnetic switching during the polarization reversal comes from the magnetoelectric coupling including the rotation of bulk anisotropy energy and the $H_{DMI}$. Besides, since BiFeO$_3$ is a G-type antiferromagnet with weak magnetization, the demagnetization energy is negligible compared to that of a ferromagnet. Hence, the damping precession is suppressed, and the switching speed becomes faster. If we look at the trajectories of **N** switching, it is interesting that the switching trajectories of **N** only lies on the *xy* plane and thus, **N** switches much faster with less precession or oscillation compared to the magnetic movement of a ferromagnet.

Next, we look at the simulation results when the DM interaction is calculated using a spin Hamiltonian instead of a magnetic field (Figure 5). It can be seen that the switching curves obtained from the numerical micromagnetic model with the DMI spin Hamiltonian and OOMMF using an effective magnetic field representing the DM field match well. Our findings contradict the results from the first principle calculations in [22], which indicates that $M_c$ switches but $N$ does not after polarization reversal. However, both of these scenarios are consistent with literature[1] in BiFeO$_3$/CoFe heterojunction which only track the easy axis of BiFeO$_3$ remains unchanged after polarization switching. In fact, in our simulations we also see cases in which $M_c$ switches and $N$ does not if the polarization switches very fast as will be shown in the later discussion. From our numerical micromagnetic model, we can obtain the spatial distribution of the effective magnetic field in each sublattice. In Figure 6 one can see that $H_{ex,ij}$, $H_{ani,bulk}$, and $H_{dem,i}$ are staggered fields, while $H_{DMI}$ is a uniform field. The reason for a uniform DM field is because the magnetization and $D_{i,j}$ are both staggered vectors. Hence, we believe that the cross product of them can be approximated as an effective magnetic field as we have done in our OOMMF model. For an antiferromagnet, a staggered field creates a precessing torque on Neel vector while a homogeneous field cants the magnetic moment without reorienting the Neel vector if the magnetic field does not reach spin-flop transition field[32], which is consistent with our results.

We then evaluate the theoretical antiferromagnetic switching time of a single-domain BiFeO$_3$ thin film from switching time of $N$ for various hypothetical polarization switching times. In previous experiments, it has been reported that increasing the magnitude of the electric field in (100) BiFeO$_3$ thin film improved the polarization switching time from second to microsecond[16]. However, those experiments involved very large samples size (~area of 750 $\mu m^2$ and thickness of 300 nm) and the parasitics of the sensing circuits were quite dominant[33]. We analyze the switching time of $N$ with the varying polarization switching time $T_{FE}$ from 10 ps to 1 ns in Figure 7. The results show that the switching time of $N$ is proportional to $T_{FE}$, and in general, is longer than $T_{FE}$. In addition, when the polarization switching time is as fast as 10 to 20 ps, the spin vectors in the antiferromagnet cannot respond and $N$ remains non-switched whereas $M_c$ switches, which corresponds to a switching failure of $N$; and if $T_{FE}$ is larger than 30 ps, $N$ switches but $M_c$ does not switch. Therefore, 30 ps is theoretically calculated as the lower limit of the antiferromagnetic switching time of a BiFeO$_3$ thin film device when $K_{AFM}$ is -1.75 $\times 10^4$ J/m$^3$. We can also roughly calculate this minimum time of magnetization reorientation, or called magnetic relaxation time, using $\frac{1}{f} = \frac{2\pi}{\gamma} \frac{1}{H_{eff}} \approx 36\ ps$ where $f$ is the ferromagnetic resonance frequency, and $H_{eff}$ is the effective applied field, which is $H_{ani,bulk}$ in the case of BiFeO$_3$. So when $T_{FE}$ is shorter than the magnetic relaxation time, $N$ cannot switch but the coupling between $P$, $N$, and $M_c$ then force $M_c$ to switch 180$^0$ after polarization reversal.

It is also important to know the critical parameters that affect the switching dynamics of BiFeO$_3$. To check the sensitivity of parameters, such as $J_{AFM}$, $K_{AFM}$, and $K_{EPI}$, the magnitude of these parameters are varied in the switching dynamics of an antiferromagnet. We find that when the exchange coupling field becomes smaller than $-2.6 \times 10^{-13}$ J/m, the switching trajectories of $M_c$ becomes highly oscillatory. This is because the deviation of spin vectors from their preferred axis increases under a weak exchange coupling as shown in Figure 8. In contrast to the case when the coupling is strong, the switching of $N$ fails and the switching time of $M_c$ increases because of the characteristics of oscillatory switching. Regarding to $K_{AFM}$ and $K_{EPI}$, one needs to note that they both affect the magnitude of the energy barrier. However, $K_{AFM}$ also affects the driving force for magnetization switching because $K_{AFM}$ originates from the DMI and is proportional to the polarization and the electric field. Thus, increasing $K_{AFM}$ increase the effective field and the energy barrier of the AFM thus implies a higher switching success rate of **N** under thermal noise, a shorter minimum input pulse width, and a shorter switching time because of a shorter period of oscillation. For the compressive constraint, increasing $K_{EPI}$ affect the crystal structure of $BiFeO_3$ to be more tetragonal-like and magnetic moments to lie in-plane. To have a successful and faster Neel vector switching, it is important to increase the magnitude of $K_{AFM}$ and $K_{EPI}$ for sufficient energy barrier to alleviate the thermal noise effect and also reduce the relaxation time of $BiFeO_3$.

**Conclusion**

We have analyzed the switching dynamics of a single-domain $BiFeO_3$ thin film by solving LK and LLG equations, simultaneously. Our results show that $BiFeO_3$ as a G-type antiferromagnet has staggered spin vectors thus staggered DM vector, which creates a weak magnetization by tilting spin vectors unidirectionally. From the analysis of the energy landscape, we also demonstrate that the preferred axis of the magnetic moment in $BiFeO_3$ is determined by both the bulk DMI energy that couple to polarization and the epitaxial strain that comes from substrate. We then show for the first time that **N** rotates $180^0$ while $\boldsymbol{M_c}$ remains unchanged by rotating polarization $180^0$ and verify the results by solving the effective equation of motion for the Neel vector in the LG theory. The driving force of the magnetic switching is due to the magnetoelectric coupling such that the easy plane state and $\boldsymbol{H_{DMI}}$ rotate along with the polarization. By checking the sensitivity of the parameters, we find that the probability of switching of **N** depends not only by the anisotropy energy barrier but also on the exchange coupling field in $BiFeO_3$. We further calculate the lower limit of the switching time of $BiFeO_3$ to be around 30 ps assuming the polarization can be switched as fast. **N** cannot be switched if the polarization switches faster than 30 ps. The results shown in this paper are of importance to justify $BiFeO_3$ as a promising material for fast voltage-controlled devices such as magnetic-random-access memory and ME beyond-CMOS logic.

*Table 1 Simulation parameter in the ferroelectric dynamics model* [23]

| Variable | Value | Units (SI) |
|---|---|---|
| $P_s$ | 0.8 | $Cm^{-2}$ |
| $\alpha_1$ | 4.9(T-1103) ×10⁵ = -3.935×10⁸ when T=300 K | $C^{-2}m^2N$ |
| $\alpha_{11}$ | 6×10⁸ | $C^{-4}m^6N$ |
| $\alpha_{12}$ | -1×10⁶ | $C^{-4}m^6N$ |
| $\gamma_{FE}$ | 5×10⁻³ | msec/F |
| $K_{Strain}$ | 6×10⁸ | N/m² |
| $\epsilon_r$ | 54 | – |
| $E_{ext}$ | 3×10⁸ | V/m |

*Table 2 Simulation parameters in the magnetic dynamics model*

| Variable | Value | Units (SI) |
|---|---|---|
| $M_s$ | 4.26×10⁵ | A/m |
| $K_{AFM}$ | -1.75 ×10⁴ | J/m³ |
| $K_{EPI}$ | -1.75 ×10⁴ | J/m³ |
| $A_{AFM}$ | -2.6×10⁻¹² | J/m |
| $H_{DMI}$ | 1000 | Oe |
| $\alpha$ | 0.01 | - |
| $\gamma$ | 2.21 ×10⁵ | $(A/m)^{-1}s^{-1}$ |

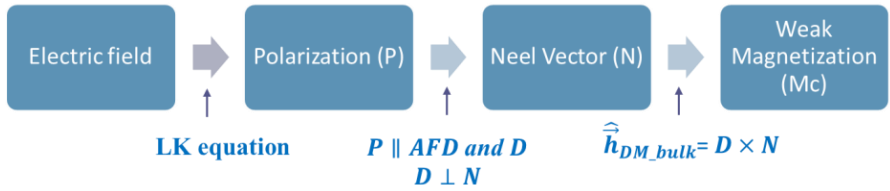

*Figure 1 Strategies of the ferroelectric and micromagnetic simulation model of BiFeO₃*

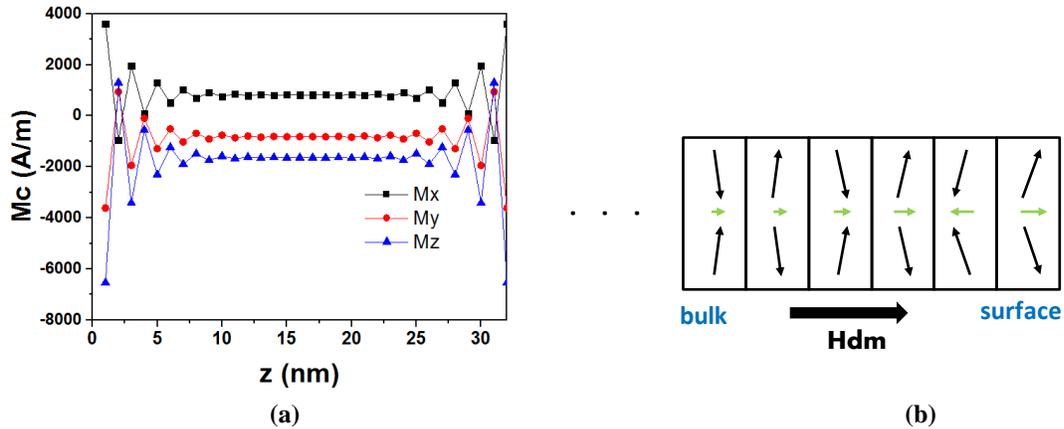

*Figure 2 (a) $M_c$ in 32 nm AFM with $H_{DMI}$=1000 Oe. (b) Schematic of magnetic moments inside a 1D antiferromagnet array under weak DM field. The black arrows represent the spin vectors and the green arrows represent the direction of $M_c$.*

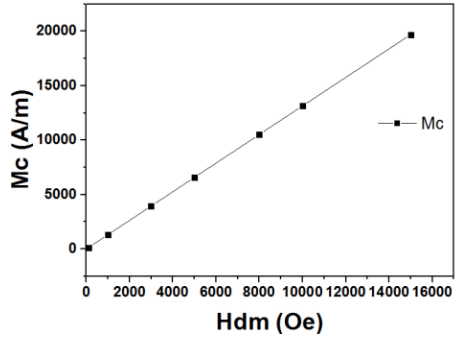

Figure 3 A 32 nm BiFeO$_3$ thin film, $K_{AFM} = K_{EPI} = -1.75 \times 10^4$ J/m$^3$

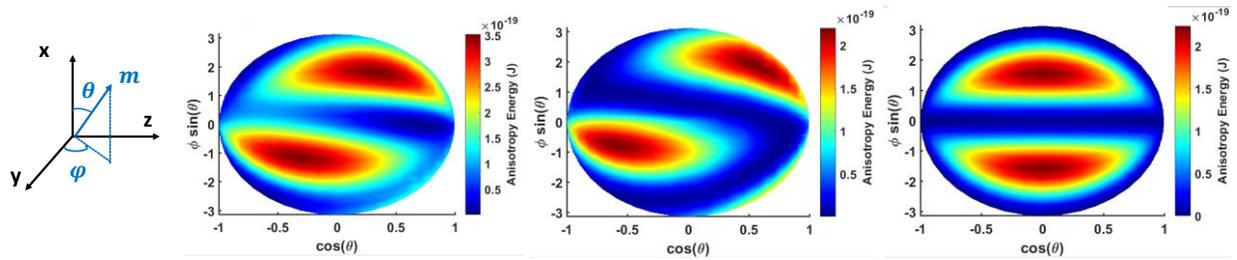

Figure 4 Energy landscape of BiFeO$_3$ thin film when (a). $K_{AFM}=K_{EPI}=-1.75 \times 10^4$ J/m$^3$ (b). $K_{AFM} = -1.75 \times 10^4$ J/m$^3$ and (c). $K_{EPI}=-1.75 \times 10^4$ J/m$^3$

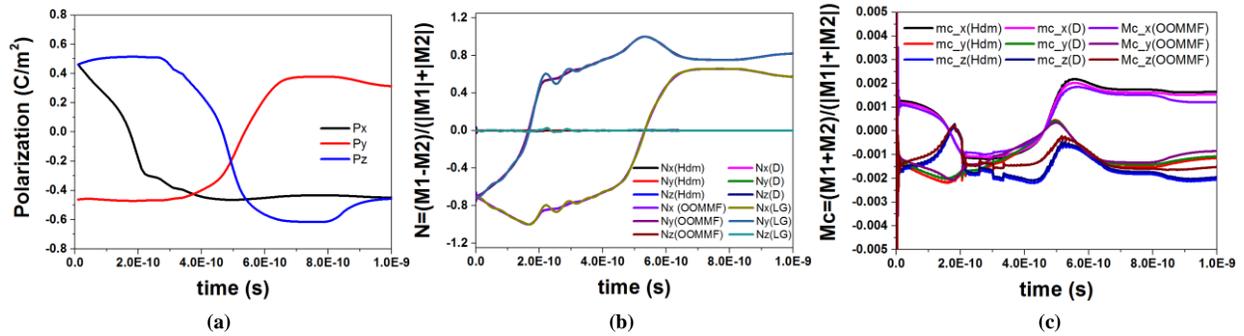

(a)    (b)    (c)

Figure 5 Switching dynamics of (a). polarization, (b). Neel vector, and (c). weak magnetization in 10 nm BiFeO$_3$ thin film with $K_{AFM}=K_{EPI}=-1 \times 10^6$ J/m$^3$ and $H_{DMI} = 1000$ Oe simulated by OOMMF, MATLAB where DMI is an effective magnetic field (Hdm) or from spin Hamiltonian (D), and LG theory.

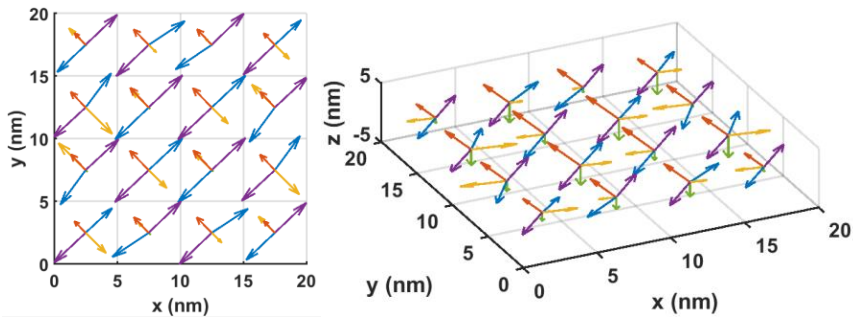

Figure 6 Spatial distribution of Heff including $H_{ex}(violet)$, $H_{ani,bulk}(yellow)$, $H_{ani\_epi}(green)$, $H_{dem}(blue)$, and $H_{DMI}(red)$ in AFM with z=1 nm.

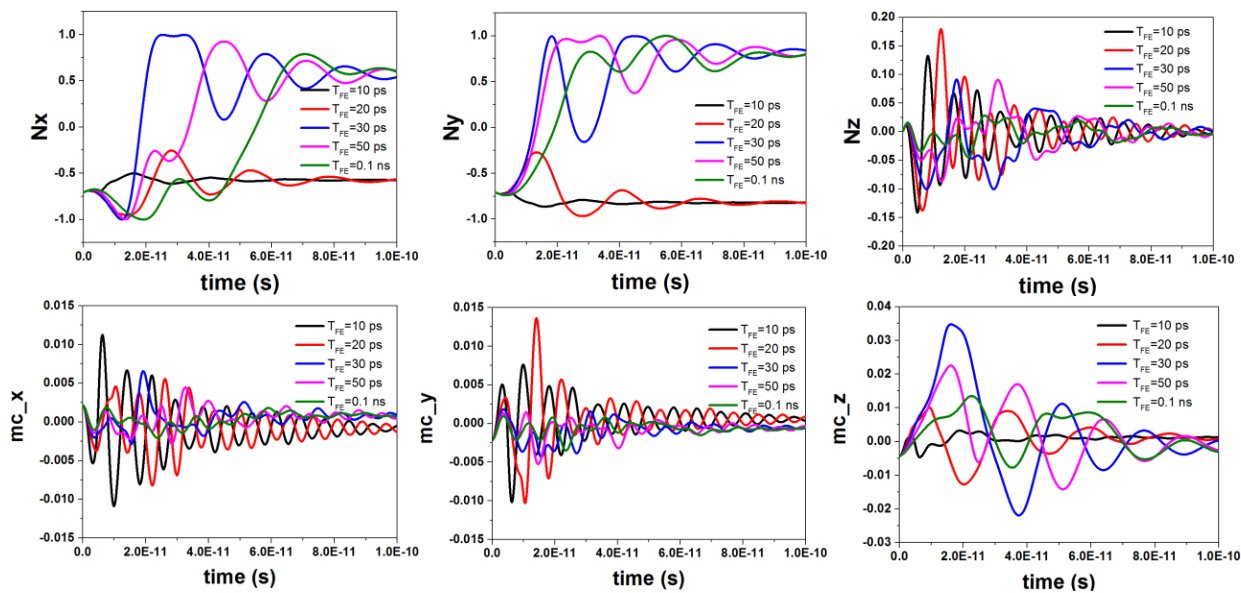

*Figure 7 Switching curves of BiFeO$_3$ thin film with varying T$_{FE}$*

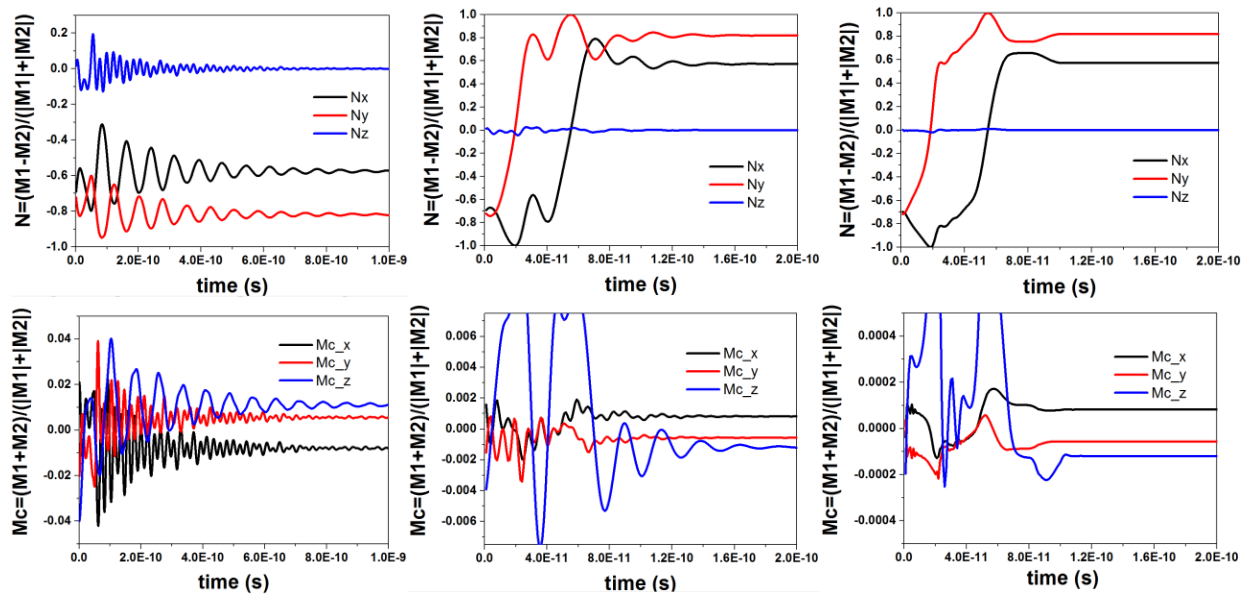

*Figure 8 The switching dynamics of Neel vector and M$_c$ of the BiFeO$_3$ thin film with J$_{AFM}$=-0.26 pJ/m, -2.6 pJ/m, and -26 pJ/m (left to right)*